\documentclass[12pt]{article}
\usepackage{graphicx,cite}

\def\baselinestretch{1.5}

\renewcommand{\thefootnote}{\fnsymbol{footnote}}

\renewcommand{\title}[1]{\bigskip\bigskip\Large\bf #1\bigskip\bigskip\\}
\renewcommand{\author}[1]{\large\rm #1\\ \bigskip}
\newcommand{\address}[1]{{\normalsize\it #1\\}\bigskip}

\begin{document}


\vglue .3 cm

\begin{center}
\title{Thermodynamic Bethe Ansatz for the subleading magnetic
	perturbation of the tricritical Ising model}

\author{R. M. Ellem$^{1,}$\footnote[2]{email: 
		{\tt rme105@rsphysse.anu.edu.au}},
	V. V. Bazhanov$^{1,2,}$\footnote[3]{email:
		{\tt Vladimir.Bazhanov@anu.edu.au}}}

\address{$^{1}$Department of Theoretical Physics, RSPhysSE, IAS,\\
	Australian National University, Canberra, ACT 0200, Australia.}
\address{$^{2}$Centre for Mathematics and its Applications, IAS,\\
	Australian National University, Canberra, ACT 0200, Australia.}
\end{center}

\setcounter{footnote}{0}
\vspace{5mm}

\begin{abstract}
We give further support to Smirnov's conjecture on the exact 
kink S-matrix for the massive Quantum Field Theory
describing the integrable perturbation of the $c=0.7$ minimal Conformal
Field theory  (known to
describe the tri-critical Ising model) by the  operator $\phi_{2,1}$. 
This operator has conformal dimensions 
$(\frac{7}{16},\frac{7}{16})$ and is identified with the subleading
magnetic operator of the tri-critical Ising model.
In this paper we apply the Thermodynamic Bethe Ansatz (TBA) approach
to the kink scattering theory by explicitly utilising its relationship 
with the solvable lattice hard hexagon model. Analytically examining 
the ultraviolet scaling limit we recover the expected central charge
$c=0.7$ of the tri-critical Ising model. We also compare numerical 
values for the ground state energy of the finite size system obtained
from the TBA equations with the results obtained by the Truncated Conformal 
Space Approach and Conformal Perturbation Theory.
\end{abstract}

\newpage

\renewcommand{\thefootnote}{\arabic{footnote}}
\section{Introduction}
\setcounter{equation}{0}

Since the work \cite{ift} of A. B. Zamolodchikov, it is known that 
certain perturbations of conformal field theories (CFT's) lead to 
completely integrable models of massive quantum field theory (QFT).
In this paper we consider the relevant perturbation of the $c=0.7$
minimal CFT   \cite{bpz} (known to describe the tri-critical Ising
model (TIM) \cite{fqs}) by the operator  $\phi_{2,1}$. This operator 
has conformal dimensions $(\Delta,\bar{\Delta}) = 
(\frac{7}{16},\frac{7}{16})$ and is identified with the subleading 
magnetic operator of the TIM. The action of the perturbed 
tri-critical Ising model (PTIM) can be written as
\begin{equation}
{\cal A}_{PTIM} = {\cal A}_{TIM} + g \int \phi_{2,1}(x) d^{2}x 
\label{eq:action}
\end{equation}
where ${\cal A}_{TIM}$ represents the action of the $c=0.7$ minimal 
CFT and $g$ is a coupling constant, of dimension $( \mbox{mass})^{
\frac{9}{8}}$, which describes the strength of the perturbation. It 
was shown in \cite{ift} that the action (\ref{eq:action}) describes 
a completely integrable QFT in the sense that it possesses an infinite 
number of non-trivial local integrals of motion. 

This model has been already studied by several authors 
\cite{smirnov,lassig,zam0,ckm,klamelz1}. 
Their results (with some violations in chronological order) are 
summarised below. The truncated conformal space approach (TCSA) 
calculations by L\"assig, Mussardo and Cardy \cite{lassig} showed that 
the theory (\ref{eq:action}) has two degenerate vacuum  states 
(labelled here by 0 and 1) which correspond to the minima of the 
asymmetric double well Landau-Ginsburg potential. It was suggested 
\cite{lassig} that the particle spectrum consists of a triplet 
of fundamental kink states\footnote{ Note that the vacuum labels 0 and 1 
of paper \cite{lassig,ckm} have been interchanged here for consistency 
with the usual convention of spin labelling in the HHM \cite{bax1}. } 
$|K_{01} \rangle$, $|K_{10} \rangle$, and $|K_{00} \rangle$
of the same mass $m$, which interpolate between these vacuum states 
(e.g. the kink $|K_{01} \rangle$ interpolates between the vacuum  
states $0$ and $1$, etc.). The fourth possible kink state $|K_{11} 
\rangle$ is not present due to the absence of the $Z_2$-symmetry 
between the vacuum states which is broken by the perturbation term 
in (\ref{eq:action}). This leads to a  restriction on the allowable 
adjacent vacuum states, which (as noticed by Zamolodchikov in 
\cite{zam0}) is equivalent to the restriction imposed on adjacent 
spin states in the solvable lattice hard hexagon model (HHM) 
\cite{bax2}. Therefore, assuming the above particle spectrum is exact, 
one can construct  the kink-kink $S$-matrix from the associated  
Boltzmann weights of the (critical) HHM, since they satisfy the same 
Yang-Baxter  equation. The normalisation of the $S$-matrix  is then 
determined \cite{klamelz1} by imposing unitarity, crossing symmetry 
and bootstrap requirements.  The resulting $S$-matrix is given 
explicitly in the next Section.  Alternatively, (and, in fact, earlier)
Smirnov \cite{smirnov} has conjectured  $S$-matrices for all 
$\phi_{1,2}$ and $\phi_{2,1}$ perturbed minimal models expressing them 
in terms of certain (appropriately normalised) RSOS projections of the 
$A_{2}^{(2)}$ R-matrix of the Izergin-Korepin model \cite{izerkor}. 
In particular, for the special case of the PTIM we are interested in 
here his result leads exactly \cite{ckm} to the same particle structure 
and $S$-matrix as discussed above. It must, however, be emphasised that 
both of these approaches are still based on conjectures and as such 
require further verification. 

In this paper we apply the thermodynamic Bethe Ansatz (TBA) approach 
\cite{yang2,zam1} to the conjectured $S$-matrix of the PTIM and 
calculate the ground state energy $E(R)$ in the finite-size geometry 
with the spatial coordinate compactified on a circle of circumference 
$R$. The calculations utilise the above relationship between the PTIM 
and the lattice hard hexagon model allowing the use of known results 
from the analytic Bethe-Ansatz solution \cite{bazresh} of the latter.
The ground state energy $E(R)$ is expressed in terms of the solution of 
the TBA integral equation of an apparently new type. We show that in 
the ``ultraviolet'' limit $R\rightarrow0$ the ground state energy scales 
as
\begin{equation}
E(R) \sim - \frac{\pi c}{6 R}  \label{eq:ec}
\end{equation}
with $c=0.7$, exactly as one expects from (\ref{eq:action}).  
We also solve the TBA equations and evaluate the ground state energy 
$E(R)$ numerically. The results are in a good agreement with those 
obtained by the TCSA \cite{lassig,ckm} and from the Conformal 
Perturbation Theory (CPT) \cite{fateev}.

\section{The Bethe-Yang equations}
\label{sec-scatt}
\setcounter{equation}{0}

As mentioned above the kink-kink $S$-matrix in the PTIM is expressed in 
terms of the Boltzmann weights of the critical HHM. The latter is an 
``interaction-round-a-face'' model on the square lattice. With a 
suitable normalisation convenient for our purposes the Boltzmann weights 
of this model can be written in the form \cite{bax1}
\begin{eqnarray}
W \! \! \left( \left. \begin{array}{cc} 0 & 0 \\
0 & 0 \end{array}  \right| u \right) & = & \frac{\sin \mu \,\sin (2\mu 
+ u)}{\sin 2\mu\, \sin (\mu-u)} \\
W \! \! \left( \left. \begin{array}{cc} 0 & 0 \\
1 & 0 \end{array}  \right| u \right) & = & W \! \! \left( \left.
\begin{array}{cc} 0 & 1 \\ 0 & 0 \end{array}  \right| u \right) = 
\left[\frac{\sin\mu}{\sin2\mu}\right]^{\frac{1}{2} }
{\sin u\over \sin(\mu - u)}\\
W \! \! \left( \left. \begin{array}{cc} 1 & 0 \\
0 & 0 \end{array}  \right| u \right) & = & W \! \! \left( \left.
\begin{array}{cc} 0 & 0 \\ 0 & 1 \end{array}  \right| u \right) = 1\\
W \! \! \left( \left. \begin{array}{cc} 0 & 1 \\
1 & 0 \end{array}  \right| u \right) & = &{\sin \mu \over \sin 2\mu}
 \frac{ \sin (2\mu -u)}{\sin(\mu-u)} \\
W \! \! \left( \left. \begin{array}{cc} 1 & 0 \\
0 & 1 \end{array}  \right| u \right) & = & \frac{ \sin (\mu +u)}{\sin
(\mu-u)}  
\end{eqnarray}
where $u$ is the spectral parameter and $\mu = \pi/5$. Note that this
model is also solvable for other values of $\mu$ such that $5\mu / 
\pi={\rm integer}$, where it corresponds to some non-unitary QFT (see 
Sect. 6 below). For the above normalisation of the Boltzmann weights 
the partition function per site of the model in the limit of the 
infinite lattice is given by \cite{bax1}
\begin{equation}
\kappa(u,\mu)=\exp\left(\int_{-\infty}^{\infty} {\sinh(\pi-\mu)x \sinh
2 u x\over 2x\,\sinh\pi x \,\cosh\mu x}\,dx\right),
\qquad 0\le{\rm Re}\, u\le\mu.\label{part}
\end{equation}
In particular for $\mu = \pi/5$ it reads  
\begin{equation}
\kappa(u,{\pi/5})={\sin(\pi/5+u)\sin(2\pi/5-u)\over\sin(\pi/5-u)
\sin(2\pi/5+u)}.  \label{part1}
\end{equation}
The kink-kink $S$-matrix proposed by Smirnov \cite{smirnov} can then be 
written as\footnote{ We follow \cite{ckm} for the spin labelling 
convention in the elements of the $S$-matrix, except for the 
aforementioned interchange of vacuum labels $0$ and $1$.}
\begin{equation}
S_{\alpha \beta}^{\gamma \delta} ( \theta ) = \left( \frac{\rho_{
\gamma} \rho_{\delta}}{\rho_{\alpha} \rho_{\beta}} \right)^{- 
\frac{\theta}{2 \pi i}} R ( \theta )\,\, W \! \! \left( \left. 
\begin{array}{cc} \alpha & \delta \\ \gamma & \beta \end{array} 
\right| \lambda \theta \right), \; \; \; \; \alpha,\beta,\gamma,\delta 
\in \{ 0,1 \}  \label{eq:smat}
\end{equation}
where $\theta$ is the rapidity variable, $\rho_0=1,\ \rho_1=2\cos\mu$,
while the other parameters are 
\begin{equation}
\mu={\pi\over5},\qquad \lambda={9i\over5}\ . \label{eq:lambda}
\end{equation}
The normalisation factor in (\ref{eq:smat}) is given by
\begin{equation}
R( \theta ) = \kappa(\lambda\theta,\mu)^{-1}\, F_{CCD}(\theta) 
\label{eq:normal}
\end{equation}
where  $F_{CDD}(\theta)$ denotes the CDD factor
\begin{equation}
F_{CDD}(\theta)=-F_{- \frac{1}{9}}\, ( \theta ) F_{\frac{2}{ 9}} 
(\theta ) ,\qquad F_{\alpha} ( \theta ) = 
{\sinh\theta+i\sin\alpha\pi\over \sinh\theta-i\sin\alpha\pi}.
\label{eq:fcdd}
\end{equation}
It was shown \cite{smirnov,ckm,klamelz1} that the $S$-matrix 
(\ref{eq:smat}) satisfies all the unitarity, crossing symmetry and 
bootstrap requirements.

Consider the state of $N$ kinks distributed along a large spatial circle 
of the length $L$ so that their average mutual distances are much greater 
than the correlation length $\xi = 1/m$ of the system, where $m$ is the 
kink mass. Since the kink scattering processes are of the factorised type, 
this state can be described in terms of the Bethe wave function 
\cite{yang2},\cite{zam2}
\begin{equation}
\Psi ( \theta_{1}, \ldots , \theta_{N} ) = \sum_{\{\alpha\}}
\psi^{\{\alpha\}} | K_{\alpha_{1} \alpha_{2}} ( \theta_{1} )
K_{\alpha_{2} \alpha_{3}} ( \theta_{2} ) \ldots K_{\alpha_{N}
\alpha_{1}} ( \theta_{N} ) \rangle  \label{eq:wavefn}
\end{equation}
where $\psi^{\{\alpha\}}$ denotes the ``colour'' wave function depending 
on the set of interkink vacuum labels $\{\alpha\}=\{\alpha_{1},\alpha_{2}, 
\ldots,\alpha_{N}\}$, $\alpha_i=0,1$, along the circle. Note that the 
absence of the kink $K_{11}$ in the theory implies that the sum in 
(\ref{eq:wavefn}) is taken over only those sequences of $\{\alpha\}$ 
which do not contain two consecutive $\alpha$'s  equal to $1$ in any place.

Consistency of the imposed periodic boundary conditions requires that
\begin{equation}
e^{imL \sinh  \theta_{k}} \sum_{\{\alpha\}} T(\theta_k;\theta_{1},\ldots
,\theta_{N} )_{\left\{ \alpha \right\}}^{\left\{ \beta \right\}}
\psi^{\{\alpha\}} = -\psi^{\{\beta\}} , \; \; \; \; \; \; \; k=1, \ldots
,N   \label{eq:trwv}
\end{equation}
where
\begin{equation}
T(\theta;\theta_{1},\ldots ,\theta_{N} )_{\left\{ \alpha
\right\}}^{\left\{ \beta \right\}} = \prod_{i=1}^{N} S_{\beta_{i}
\alpha_{i+1}}^{\alpha_{i} \beta_{i+1}} ( \theta - \theta_{i} )  
\label{eq:trm}
\end{equation}
can be interpreted as the transfer matrix of a two-dimensional lattice 
model acting in the space of colour wave functions $\psi^{\{\alpha\}}$.  
The equations (\ref{eq:trwv}) result in constraints on the set of the 
kink rapidities $\theta_1,\theta_2,\ldots,\theta_N$ which  are known as 
the Bethe-Yang (BY) equations
\begin{equation}
e^{imL \sinh  \theta_{k}} \Lambda(\theta_{k};\theta_{1},\ldots ,
\theta_{N} ) = -1 , \; \; \; \; \; \; \; k=1, \ldots ,N \label{eq:betye}
\end{equation}
where  $\Lambda(\theta;\theta_{1}, \ldots ,\theta_{N} )$ are the 
eigenvalues of the  transfer matrix (\ref{eq:trm}). From the definition 
(\ref{eq:smat}) it is obvious that (\ref{eq:trm}) is just the 
transfer matrix of the (inhomogeneous) critical HHM. The eigenvalues of 
the latter were found in \cite{bazresh} by solving the  transfer matrix 
functional equation. Using this result one obtains
\begin{equation}
\Lambda ( \theta;\theta_1,\ldots,\theta_N ) = \Lambda_{HHM} ( \lambda
 \theta;\theta_1,\ldots,\theta_N ) \ \prod_{j=1}^{N} R( \theta - 
\theta_{j} ) 
\end{equation}
where
$$
\Lambda_{HHM} (u;\theta_{1},\ldots ,\theta_{N})  =  \omega 
 \frac{Q(u+ \mu )}{Q(u)} + \omega^{-1} f(u) \frac{Q(u- \mu )}{Q(u)}  
$$
\begin{equation}
f(u) \! \!  =  \! \! \prod_{j=1}^{N} \frac{\sin (u-\lambda \theta_{j})}{
\sin (u-\lambda \theta_j-\mu)} ,\qquad Q(u) \! \!  =  \! \! \prod_{k=1}^{
N/2} \sin \left( u- \frac{\mu}{2} -\lambda\alpha_k \right)  
\label{eq:eigen}
\end{equation}
$$
\omega^{5}  = -(-1)^{N/2}, \qquad \omega \neq \pm 1
$$
and where $N$ is assumed to be even, and the trivial difference in the 
normalisation of the transfer matrix (\ref{eq:trm}) with that of ref. 
\cite{bazresh} has been taken into account. The (complex) numbers 
$\alpha_{k}$, $k=1, \ldots , N/2$, are determined by the following 
Bethe-Ansatz equations (BAE)
\begin{equation}
\omega^{-2} \prod_{j=1}^{N}S_1(\alpha_k-\theta_j) = - \prod_{l=1}^{n} 
S_2(\alpha_k-\alpha_l) \label{eq:nbae}
\end{equation}
where 
\begin{equation}
n={N\over2}
\end{equation}
and 
\begin{equation}
S_j(\theta)= {\sin(\lambda \theta +\mu j/2)\over\sin(\lambda \theta - \mu 
j/2)}. \label{eq:sfunc}
\end{equation}
Note that the equations (\ref{eq:nbae}) can be considered as the BAE for 
the $n=N/2$ up-arrow sector of the inhomogeneous six-vertex model 
\cite{baxbook} with ``twisted'' boundary conditions. The factor 
$\omega^{-2}$ in the RHS is then interpreted as the ``twisting factor'' 
determined by the value of the (horizontal) field applied to the six-vertex 
model.

\section{The Thermodynamic Limit}
\setcounter{equation}{0}

According to the standard approach of the Thermodynamic Bethe Ansatz (e.g. 
\cite{zam1},\cite{mus1}) the BY equations (\ref{eq:betye}) allow one to 
determine the spectral density of states of the infinite size system and 
hence to compute the free energy $f_0(mR)$ of this system at a finite 
temperature $1/R$. This free energy is then reinterpreted as the Casimir 
part of the ground state energy $E(R)$ of the finite-size system on a 
circle of the circumference $R$
\begin{equation}
E(R) -{\rm (bulk \  term)}= Rf_{0}(mR)   \label{eq:erf}
\end{equation}
where $f_{0}(mR)$ is the infinite-size free energy per unit length and the 
linear  in $R$ bulk energy term is known explicitly (see eq.(\ref{eq:eofr}), 
below).

In the thermodynamic limit the number of kinks $N$ and the system size $L$ 
simultaneously approach infinity. From earlier studies of the six vertex 
model it is well known \cite{taka} that for $N\rightarrow\infty$ the roots 
$\alpha_k$ which solve the BAE (\ref{eq:nbae}) approach certain asymptotic 
patterns in the complex plane which can viewed as collections of ``strings''. 
Each string is a set of roots  with the same real part, symmetric with 
respect to the real axis of $\alpha$ and equally spaced along the imaginary 
axis. The number of roots in a string is called the ``length'' of the string.
In general the string spectrum of (\ref{eq:nbae}) may be very complicated 
\cite{taka} since it heavily depends on the arithmetic properties of the 
constant $\mu$. For the case under consideration the constant $\mu$ is given 
by the simple fraction $\mu=\pi/5$ and the string spectrum of (\ref{eq:nbae}) 
consists of only five different strings. These are the strings of lengths 
$\ell=1,2,3,4$ (to be referred as 1-string, 2-strings, etc.) which are centred 
at the real axis ${\Im}m\,\alpha=0$ and the ``shifted 1-string'' which is 
located on the line ${\Im}m\,\alpha={\pi/(2\lambda)}\pmod{\pi/\lambda}$ \ and 
shifted from the real axis by half of the period of the function 
$Q(\lambda\theta)$. With these strings the roots $\alpha_k$ for a typical 
solution of (\ref{eq:nbae}) are given by
\begin{equation}
\alpha^{(\ell)}_{j,m} = \beta^{(\ell)}_{j}+{\mu\over 2\lambda} (\ell+1-2m)
+ \delta(N)\pmod {\pi/\lambda} ,
\qquad m=1,\ldots,\ell,\qquad j=1,\ldots,n_\ell 
\label{eq:strng}
\end{equation}
and
\begin{equation}
\alpha^{(-)}_{j} = \beta^{(-)}_{j}+{\pi\over 2\lambda}\pmod {\pi/\lambda},
\qquad j=1,\ldots,n_-
\end{equation}
where $\beta^{(\ell)}_{j}$ and $\beta^{(-)}_j$ are (real) centres of the 
strings while $n_{\ell}$, ${\ell}=1,2,3,4$ and $n_-$ are the total numbers 
of strings  of these types. 
The correction term  $\delta(N)$ in (\ref{eq:strng}) describes deviations 
of the roots from exact string positions. For large $N$ it
vanishes 
\cite{dorfel}\footnote{Actually, the behaviour of $\delta(N)$ for large $N$ is
more complicated \cite{dorfel}. The asymptotics  $\delta(N)\simeq O(\log
N/N)$ quoted above are valid for the region ${\rm Re} |\alpha|\ll \log N$ 
used for the thermodymic limit below.}   as
$O(\log N/N)$.
In the case of the HHM we only require those
solutions of (\ref{eq:nbae}) where the total number of roots is equal to 
$n=N/2$ and hence
\begin{equation}
\sum_{\ell=1}^{4} \ell n_{\ell}+n_- = \frac{N}{2} . \label{eq:non1}
\end{equation}                
For a generic value of the twist factor $\omega$ on the unit circle, 
$|\omega|=1$, all the above strings are thermodynamically significant in 
the sense that for typical solutions of (\ref{eq:nbae}) with $n=N/2$ the 
ratios $n_\ell/N$, $\ell=1,\ldots,4$ and $n_-/N$ remain finite as 
$N\rightarrow \infty$. However, when $\omega$ approaches the specific 
values \ $\log\omega=\pm i k\pi/5$, $k=1,2,3,4$ \ as it is required in 
(\ref{eq:eigen}) and (\ref{eq:nbae}) the situation drastically changes. 
It turns out in this case that, for many of these solutions, a finite number  
of strings moves away to infinity and the condition (\ref{eq:non1}) can no 
longer be satisfied. More precisely, on the basis of numerical calculations 
it was conjectured \cite{bazresh} that the solutions of (\ref{eq:nbae}), 
where the fractions of the 4-strings $n_4/N$ and/or the shifted 1-strings 
$n_-/N$ are not vanishing as $N\rightarrow\infty$, do not match the condition 
(\ref{eq:non1}) and therefore have no relevance to the HHM.  This leaves 
only three thermodynamically significant strings of lengths $\ell=1,2,3$. 
Their total numbers are restricted by the relation
\begin{equation}
\sum_{\ell=1}^{3} \ell n_{\ell} = \frac{N}{2}+o(N) . \label{eq:non2}
\end{equation}
where the $o(N)$ term grows slower than $N$ for $N\rightarrow\infty$. 
Assuming that the roots $\alpha_k$ take their limiting string form 
(\ref{eq:strng}) one can rewrite the equation (\ref{eq:betye}) as
\begin{equation}
\omega e^{imL \sinh  \theta_{k}} \prod_{j=1}^N R(\theta_k-\theta_j)
\prod_{\ell=1}^3\prod_{j=1}^{n_\ell}
S_\ell(\theta_k-\beta_{j}^{(\ell)} )=-1+o(1),\label{eq:by1}
\end{equation}
where $S_\ell$ is defined in (\ref{eq:sfunc}) and the correction term $o(1)$ 
vanishes as $N\rightarrow\infty$. The last equation looks very similar to 
the BY equation for the diagonal scattering theory. In fact, it is very 
convenient to formally interpret the  
string centres $\beta^{(\ell)}_j$ appearing in (\ref{eq:by1}) as rapidities 
of some quasi-particles. These quasi-particles do not correspond to any 
observable particles in the asymptotic scattering states but rather reflect 
the nontrivial spin structure of the kink scattering. In the thermodynamic 
limit the rapidities of kinks and quasi-particles form dense distributions 
which can be described in terms of continuous densities. Let 
$\rho_0(\theta)$ denote the rapidity density of kink states and  
$\rho_\ell(\beta)$ denote such densities for the quasi-particles. We 
normalise these densities by the conditions 
\begin{equation}
L \int_{- \infty}^{\infty} \rho_0( \theta ) d \theta \! \!  =  \!
        \! N ;\qquad
L \int_{- \infty}^{\infty} \rho_{\ell} ( \beta ) d \beta \! \!  = 
\! \! n_{\ell} , \;\;\;\;\;\;\;\;\;\;\;\; \ell=1,2,3. \label{eq:norm}
\end{equation}
Following the standard procedure of the TBA one can now rewrite 
(\ref{eq:by1}) as an integral equation
\begin{equation}
{m \over 2\pi} \cosh\theta =\rho_0(\theta)+\widetilde{\rho}_0(\theta)- 
\sum_{k=0}^3 \int_{-\infty}^{\infty}\Psi_{0k}(\theta-\theta')\rho_k(
\theta') d\theta' \label{eq:beq1}
\end{equation}
where, as usual,  $\widetilde{\rho}_0(\theta)$ denotes the density of the
``holes'' \cite{yang2} in the kink rapidity distribution and
\begin{equation}
\Psi_{00}(\theta)={1\over 2\pi i}{\partial_\theta}\log R(\theta), \qquad
\Psi_{0j}(\theta)=\Psi_{j0}(\theta)={1\over 2\pi i}{\partial_\theta}\log
S_j(\theta),\qquad j=1,2,3. \label{eq:psifns}
\end{equation}
Similarly the BA equations (\ref{eq:nbae}) lead to,
\begin{equation}
0=\rho_j(\theta)+\widetilde{\rho}_j(\theta)- \sum_{k=0}^3 \int_{-
\infty}^{\infty} \Psi_{jk}(\theta-\theta')\rho_k(\theta') d\theta' 
\label{eq:beq2}
\end{equation}
where $\widetilde{\rho}_j(\theta)$, $j=1,2,3$, denotes the densities of 
holes in the quasiparticle rapidity distributions, $\Psi_{j0}(\theta)$ is 
given above in (\ref{eq:psifns}) and 
\begin{equation}
\Psi_{jk}(\theta)=-{1\over 2\pi i}\,{\partial_\theta}\, \sum_{m=1}^{k} 
\log \left[ S_{j-k+2m} ( \theta ) S_{j+k-2m} ( \theta ) \right],\qquad 
j,k=1,2,3.
\end{equation}
The equations (\ref{eq:beq1}) and (\ref{eq:beq2}) can be notably simplified. 
In fact, integrating the equation (\ref{eq:beq2}) with $j=3$ over $\theta$ 
from $-\infty$ to $\infty$ and using (\ref{eq:norm}) and the completeness 
relation (\ref{eq:non2}) one obtains
\begin{equation}
\int_{-\infty}^\infty \widetilde{\rho}_3(\theta)=0 .
\end{equation}
This means that the corresponding density of holes identically vanishes, 
$\widetilde{\rho}_3(\theta)\equiv0$. Therefore the equation (\ref{eq:beq2}) 
with $j=3$ can be used to exclude the density ${\rho}_3(\theta)$ from 
(\ref{eq:beq1}) and (\ref{eq:beq2}) expressing them through the remaining 
densities ${\rho}_j(\theta)$, $j=0,1,2$.  Another transformation, 
especially useful for a simplification of the integral kernel in these 
equations, consists in the following. The densities $\rho_j(\theta)$ and 
$\widetilde{\rho}_j(\theta)$ enter the integral equation (\ref{eq:beq2}) 
in a non-symmetric way. Indeed, the hole densities 
$\widetilde{\rho}_j(\theta)$ enter only the free term in (\ref{eq:beq2}) 
while the ``particle'' densities ${\rho}_j(\theta)$ enter the free term and 
also the integral term in (\ref{eq:beq2}). This arrangement can be changed. 
In particular, multiplying the equations (\ref{eq:beq1}) and (\ref{eq:beq2}) 
with a suitable matrix integral operator one can bring them to an equivalent 
form where the ``particle'' densities ${\rho}_j(\theta)$, for $j=1,2$ 
appear only in the free terms (the corresponding hole densities  will then 
enter the free and the integral terms). Performing all the transformations 
described above and redenoting the set of densities, 
$$\sigma_0(\theta)={\rho}_0 (\theta),\qquad \widetilde{\sigma}_0(\theta)=
\widetilde{\rho}_0 (\theta); \qquad \sigma_j(\theta)=\widetilde{
\rho}_j(\theta),\qquad \widetilde{\sigma}_j (\theta)={\rho}_j(\theta),
\qquad j=1,2,\qquad $$ 
one obtains
\begin{equation}
{m\over2\pi} \delta_{j,0}\,\cosh\theta= \sigma_j(\theta)+\widetilde{\sigma
}_j(\theta)+ s_j\,\ \sum_{k=0}^2 \int_{-\infty}^\infty \Phi_{j,k}(\theta-
\theta') \,\,\sigma_k(\theta') \,\, d\theta', \qquad j=0,1,2. \label{bae}
\end{equation} 
The new kernel reads
\begin{equation}
\Phi_{j,k}(\theta) =-(\delta_{j,k+1}+\delta_{j,k-1})\phi_0(\theta)-
s_0\,\delta_{j,0} \delta_{k,0} \phi_1(\theta) \label{eq:fnphi}
\end{equation}

\begin{equation}
\phi_0(\theta)={|\lambda|\over 2\mu}\, {1\over\cosh(\pi |\lambda| \theta 
/ \mu)},
\qquad \phi_1(\theta)={1\over 2\pi i}\partial_\theta \log F_{CDD}(\theta)
\end{equation}
where $F_{CDD}(\theta)$ is given by (\ref{eq:fcdd}) and the quantities 
$s_j$, $j=0,1,2$ are sign factors
\begin{equation}
s_0=-{\Im m \lambda\over|\Im m \lambda|},\qquad s_1=s_2=1. \label{sigfac}
\end{equation}
Of course, in the case considered where $\lambda$ is given by 
(\ref{eq:lambda}), the value of $s_0$ is fixed to be  $s_0=-1$. We have 
given the formula (\ref{sigfac}) just to illuminate the origin of these 
sign factors in (\ref{bae}). 

Thus, we have shown that in the thermodynamic limit the scattering state of 
the system of kinks in the PTIM is described by the three pairs of the 
rapidity and the hole densities ${\sigma}_j(\theta)$, $\widetilde{\sigma}_j
(\theta)$, $j=0,1,2$, which are constrained by the integral equation 
(\ref{bae}). In the next section we will use this result to compute the 
equilibrium free energy of the system at a finite temperature.

\section{The ground state energy of the finite size system}
\setcounter{equation}{0}

In the thermodynamic limit, the free energy for the system of kinks 
considered in the previous section is given by the functional
\begin{equation}
{\cal F} [ \sigma_{j} ( \theta ), \tilde{\sigma}_{j} (\theta )] = {\cal E} 
[ {\sigma}_0 ( \theta )] - \frac{1}{R} {\cal S} [ \sigma_{j} ( \theta ),
\tilde{\sigma}_{j} (\theta )] , \label{eq:fe}
\end{equation}
where $1/R$ acts as the temperature parameter
\begin{equation}
{\cal E} =m L \int_{-\infty}^{\infty} \cosh\theta \, \sigma_{0} ( \theta
) d \theta   , \label{eq:energy}
\end{equation}
is the energy of the system and ${\cal S}$ is the combinatorial entropy 
\cite{yang1} for the given set of the ``particle'' and ``hole'' densities 
$\sigma_j(\theta),\widetilde{\sigma}_j(\theta)$, $j=0,1,2$. 

Following the standard calculations the equilibrium free energy
\begin{equation}
f_{0}(mR) = {{\cal F}\over L}= - \frac{m}{2 \pi R} \int_{- \infty}^{\infty} 
\cosh \theta \, \log \big( 1+ e^{-\varepsilon_0(\theta)}\, \big)\,\, d 
\theta  \label{eq:minfe}
\end{equation}
is obtained by minimising the free energy functional (\ref{eq:fe}) with 
respect to the above set of densities. The pseudo-energies 
$$\varepsilon_j(\theta)=\log(\widetilde{\sigma}_j(\theta)/{\sigma}_j
(\theta))$$
are determined by the TBA integral equations
\begin{equation}
\varepsilon_{j} ( \theta ) =\delta_{j,0}\, mR \, \cosh \theta  + 
\sum_{k=0}^{2} s_k\, \int_{-\infty}^\infty \Phi_{j,k}(\theta-\theta') \log 
\big( 1+ e^{-\varepsilon_k(\theta')}\, \big)\,\, d \theta'\ \label{eq:tbae}
\end{equation}
where the kernel $\Phi_{jk}(\theta)$ and the sign factors $s_k$ are given by 
(\ref{eq:fnphi}) and (\ref{sigfac}) respectively. Combining (\ref{eq:minfe}) 
with (\ref{eq:erf}) and inserting the known bulk energy term, one 
obtains the ground state energy $E(R)$ of the finite size system defined on 
a circle of circumference $R$
\begin{equation}
E(R)=\epsilon \, R  - \frac{m}{2 \pi} \int_{- \infty}^{\infty} \cosh  \theta 
\, \log \big( 1+ e^{-\varepsilon_0(\theta)}\, \big)\,\, d \theta \ . 
\label{eq:eofr}
\end{equation}
where \cite{fateev}
\begin{equation}
\epsilon = - {\sqrt{3}\,m^{2}\over 24 \,\sin({\pi\over 18})}  \approx
        - 0.41560346108 \ldots m^{2} . 
\end{equation}

As usual, the leading asymptotics of $E(R)$ in the ultraviolet limit,
$R\rightarrow0$, can be calculated by using the well known 
``dilogarithm trick'' \cite{wiegmann}. 
In this limit the pseudo-energies $\varepsilon_j
(\theta)$ tend to finite constants $\varepsilon_{j} (0)$ inside the region 
$- \ln \left( \frac{2}{mR} \right) \ll \theta \ll \ln \left( \frac{2}{mR} 
\right)$. These constants satisfy the set of algebraic equations obtained 
by setting $R=0$ in (\ref{eq:tbae}). Bringing them into a symmetric form one gets
\begin{equation}
y_{j}^{(0)} = \prod_{k=0}^{2} \left( 1+ y_{k}^{(0)}\right)^{l_{jk}} 
\label{eq:sol1}
\end{equation}
where we have defined $y_{0}^{(0)} = \exp ( - \varepsilon_{0} (0))$ and 
$y_{j}^{(0)} = \exp ( \varepsilon_{j} (0))$, $j=1,2$. The coefficients 
$l_{jk}$ are given by
\begin{equation}
\Vert l_{jk}\Vert_{0\le j,k\le2} = \left( \begin{array}{ccc} - \frac{1}{3} 
& \frac{2}{3} & \frac{1}{3} \\ \frac{2}{3} & - \frac{1}{3} & - \frac{2}{3} 
\\ \frac{1}{3} & - \frac{2}{3} & - \frac{1}{3} \end{array} \right) .
\end{equation}
The equations (\ref{eq:sol1}) have the unique positive solution
\begin{equation}
y_{0}^{(0)} = \sqrt{2},\qquad y_{1}^{(0)} = 1, \qquad
y_{2}^{(0)} = \frac{1}{\sqrt{2}}.\label{yzero}
\end{equation}

Next, in the limit $|\theta| \rightarrow \infty$, the pseudo-energy 
$\varepsilon_{0}  (\theta) \rightarrow \infty$ while the other two 
pseudo-energies $\varepsilon_{1}(\theta), \varepsilon_{2} (\theta)$ tend to 
finite constants. These constants satisfy the set of equations
\begin{equation}
y_1^{(\infty)}=\left(1+\big(y_2^{(\infty)}\big)^{-1}\right)^{-{1\over2}},
\qquad y_2^{(\infty)}=\left(1+\big(y_1^{(\infty)}\big)^{-1}\right)^{-{
1\over2}},\qquad
\end{equation}
where we have defined $y_{j}^{(\infty)} = \exp (\varepsilon_{j} ( \infty ))$. 
These equations have the unique positive solution
\begin{equation}
y_{0}^{( \infty )} = 0,\qquad y_{1}^{( \infty )} = y_{2}^{( \infty)} = 
\frac{\sqrt{5}-1}{2}
\end{equation}
where for convenience we have defined $y_0^{( \infty)}=\exp(-\varepsilon_{0} 
(\infty ))=0$.

Following the standard calculations (see {\it e.g.} \cite{zam2}) one can 
deduce the leading $R \rightarrow 0$ asymptotics of the ground state energy 
\begin{equation}
E (R) \sim - \frac{1}{\pi R} \sum_{j=0}^{2} \left[ {\cal L} \left( f( 
y_{j}^{(\infty)}) \right) - {\cal L} \left(f( y_{j}^{(0)}) \right) \right] 
\label{eq:rzero}
\end{equation}
where
\begin{equation}
{\cal L} (x) = - \frac{1}{2} \int_{0}^{x} \left[ \frac{\ln (t)}{1-t} +
\frac{\ln (1-t)}{t} \right] dt 
\end{equation}
is the Rogers dilogarithm function and $f(x)= 1 /(1+x)$. With the simple 
dilogarithm identities
\begin{equation}
{\cal L}(x)+{\cal L}(1-x)={\cal L}(1),\qquad {\cal L}(1)={\pi^2\over6},
\qquad {\cal L} \Big({\sqrt{5}- 1\over2}\Big)= {\pi^2\over10}
\end{equation}
the equation (\ref{eq:rzero}) reduces to
\begin{equation}
E(R) \sim - \frac{7 \pi}{60 R}\label{shortd} 
\end{equation}
exactly as one expects from (\ref{eq:ec}) with the central charge $c=0.7$ 
of the tri-critical Ising CFT.  
The following few terms of the short distance expansion of $E(R)$ given by
(\ref{eq:eofr}) can easily be determined numerically. This is done in the 
next Section where we compare the TBA results  
with the those  of the Conformal Perturbation Theory (CPT)\footnote
{It should be noted that bulk term 
in (\ref{eq:eofr}) exactly compensates the next-to-leading, linear in $R$, 
term in the short distance expansion of the integral in
(\ref{eq:eofr}). In fact, the main correction term to (\ref{shortd})
is of order $O(R^{9/4})$, exactly as predicted by CPT.}.

\section{Numerical results}
\setcounter{equation}{0}

\def\baselinestretch{1.2}
\begin{table}[ht]
\centering
\begin{tabular}{|l||l|l|l|}
\hline
$\hfill mR\hfill$ & \hfil $E(R)/m$ (TBA)\hfill
 & \hfil (TCSA)\hfill & \hfil (CPT)\hfill \\ \hline \hline
$0.00001$ & $-36651.91429205$ & $-36651.914292$ & $-36651.9142920555$ \\
$0.000025$ & $-14660.7657173015$ & $-14660.765717$ & $-14660.7657173015$ \\
$0.00005$ & $-7330.38285968217$ & $-7330.382860$ & $-7330.38285968217$ \\
$0.000075$ & $-4886.92190775210$ & $-4886.921907$ & $-4886.92190775210$ \\
$0.0001$ & $-3665.19143229428$ & $-3665.191432$ & $-3665.19143229428$ \\
$0.00025$ & $-1466.07658143979$ & $-1466.076578$ & $-1466.07658143979$ \\
$0.0005$ & $-733.038309061764$ & $-733.038300$ & $-733.038309061764$ \\
$0.00075$ & $-488.692229111026$ & $-488.692214$ & $-488.692229111026$ \\
$0.001$ & $-366.519198155448$ & $-366.519177$ & $-366.519198155448$ \\
$0.0025$ & $-146.607830808513$ & $-146.607765$ & $-146.607830808515$ \\
$0.005$ & $-73.3042415739432$ & $-73.304084$ & $-73.3042415739653$ \\
$0.0075$ & $-48.8699046289368$ & $-48.869643$ & $-48.8699046290281$ \\
$0.01$ & $-36.6528965534080$ & $-36.652522$ & $-36.6528965536577$ \\
$0.025$ & $-14.6638535325716$ & $-14.662677$ & $-14.6638535387397$ \\
$0.05$ & $-7.33772690834948$ & $-7.334929$ & $-7.33772697813054$ \\
$0.075$ & $-4.89911302398585$ & $-4.894469$ & $-4.89911331240653$ \\
$0.1$ & $-3.68265799877659$ & $-3.676005$ & $-3.68265878811873$ \\
$0.25$ & $-1.52096720347295$ & $-1.500051$ & $-1.52098667429440$ \\
$0.5$ & $-0.86341841907386$ & $-0.813667$ & $-0.86363725527810$ \\
$0.75$ & $-0.70459594978559$ & $-0.621994$ & $-0.70548946787533$ \\
$1.0$ & $-0.67473619457720$ & $-0.556359$ & $-0.67713759026010$ \\
$2.5$ & $-1.07537196387997$ & $-0.702079$ & $-1.12306270607538$ \\
$5.0$ & $-2.08009259839356$ & $-1.182952$ & $-2.39571841186687$ \\
$10.0$ & $-4.15604421536308$ & $-1.960064$ & $-5.56031580714459$ \\
\hline
\end{tabular}
\caption{Some numerical values for the ground state energy $E(R)$ (given 
in units of mass $m$) computed from the TBA equations 
(\ref{eq:tbae}),(\ref{eq:eofr}), from the TCSA method \cite{strip}, 
and from the first three terms of the CPT short distance expansion 
(\ref{cptexp}), for different values of the dimensionless variable $mR$.}
\end{table}
\def\baselinestretch{1.5}

The TBA equations (\ref{eq:tbae}) can be solved numerically using a 
simple iterative procedure. The corresponding ground state energy $E(R)$
is then determined by numerically evaluating the integral in 
(\ref{eq:minfe}). Some resulting values of $E(R)$ in the region 
$10^{-5}<mR<10$ are given in Table~1 where they are compared to the 
corresponding values obtained from results of the Truncated Conformal 
Space Approach (TCSA) \cite{yurzam}.
\begin{figure}[ht]
\centering
\includegraphics[scale=0.66,angle=-90]{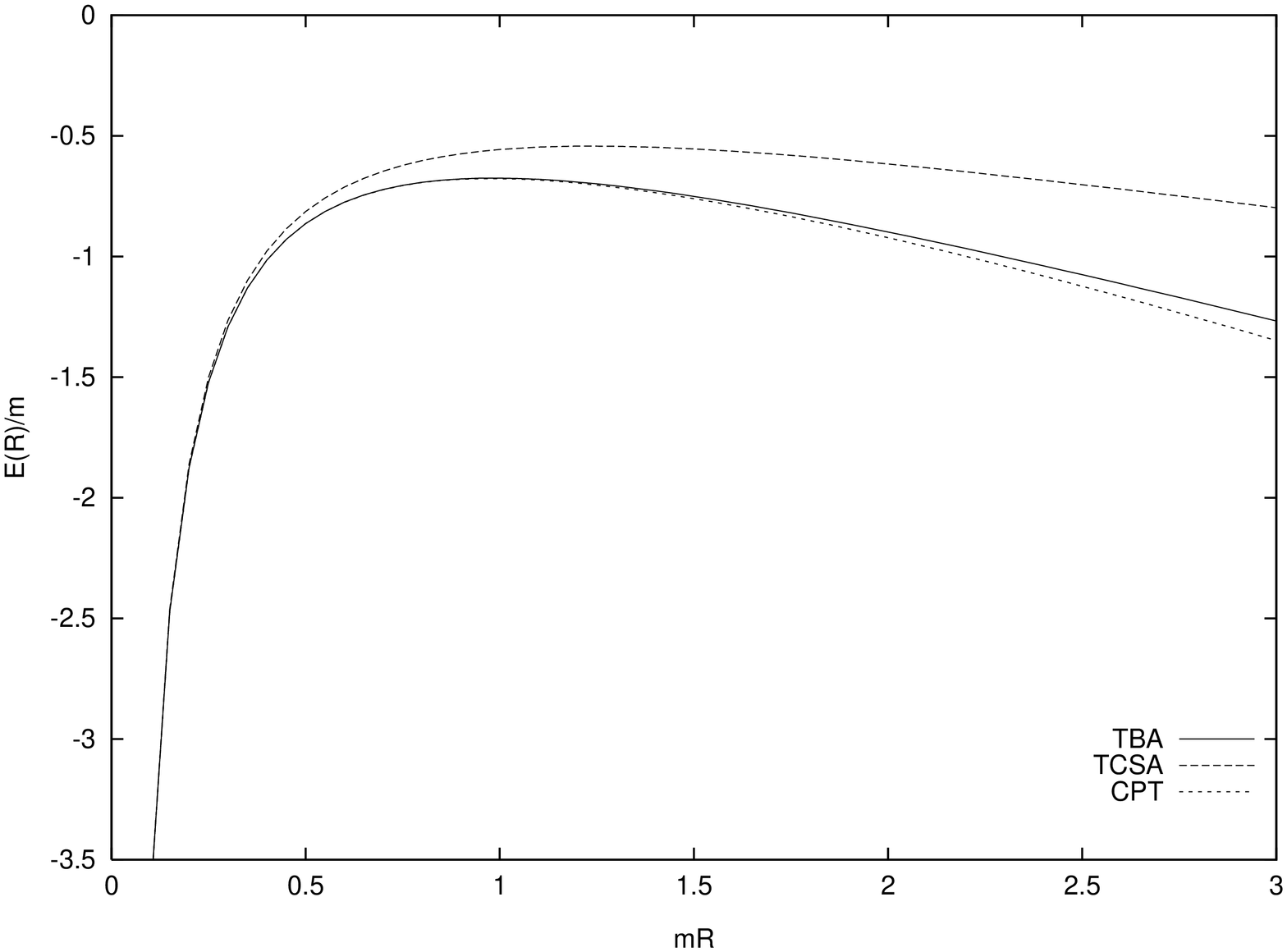}
\vspace{2.5mm}
\caption{The ground state energy $E(R)$ (given in units of mass $m$)
computed from the TBA equations (\ref{eq:tbae}),(\ref{eq:eofr}), from
the TCSA method \cite{strip}, and from the first three terms of the CPT 
short distance expansion (\ref{cptexp}), plotted against the 
dimensionless variable $mR$.}
\end{figure}

The TCSA calculations for the PTIM were performed in \cite{lassig,ckm}, 
however the required numerical data was omitted there.  We are indebted to 
Prof.~G.Mussardo who provided us with the computer program \cite{strip} 
which generates the TCSA data presented in Table~1. Note that to relate the 
TCSA data back to TBA results one has to use the exact relation 
\cite{fateev} between the kink mass $m$ and the coupling constant $g$ in 
(\ref{eq:action})
\begin{equation}
m={2^{1\over 3} \pi^{4/3} \Gamma(5/9)\over \Gamma(2/3)\Gamma(8/9)}\,
\left({2\Gamma(11/16)\over\Gamma(5/16)\Gamma(3/4)}\right)^{8/9}\,g^{8/9}
\approx 4.9277912244\ldots g^{8/9} .
\end{equation}
The difference  between the TCSA and TBA results given in Table~1 (seen 
more clearly in Fig.~1) is attributed to both the inherent inaccuracy of the 
TCSA truncation procedure (used to approximate the Hilbert space of energy 
states) which increases with $R$, and also to the fact that the dimension 
$\Delta = \frac{7}{16}$ is very close to the critical value of $\Delta = 
\frac{1}{2}$, above which divergence problems are known to occur. It is the 
latter problem which is thought to be responsible for the incorrect value of 
the bulk linear slope in the TCSA data at large $R$ (as exemplified in 
Figure~1). We refer the reader to the original paper \cite{lassig} where 
these problems are more thoroughly discussed.

In Table~1 (and Fig.~1) we also compare our numerical results for 
$E(R)$ with the first three terms of the short distance expansion 
\begin{equation}
E(R) = - \frac{\pi}{6R} \left[ {7\over 10} + \sum_{k=2}^{\infty} b_{k} 
(-\pi g)^{k} \left( \frac{R}{2 \pi} \right)^{\frac{9 k}{8}} \right]
.\label{cptexp} 
\end{equation}
obtained from Conformal Perturbation Theory (CPT)
\cite{zam1,zam2,klamelz2}.
In our case the coefficients $b_{2}$ and $b_{3}$ read explicitly  
\cite{fateev}
\begin{equation}
b_{2}  =  12 \frac{\Gamma^{2} \left( \frac{7}{16} \right) \Gamma \left( 
\frac{1}{8} \right)}{\Gamma^{2} \left( \frac{9}{16} \right) \Gamma \left(
\frac{7}{8} \right)} \approx 135.9255988883 \ldots,
\qquad b_{3}  =  0 .
\end{equation}
Considering these two coefficients as fixed and using the TBA results it is 
not difficult to make a numerical fit for a few following coefficients in 
(\ref{cptexp}). In particular, we found that 
\begin{equation}
b_4 \approx -250.3\ldots \; , \qquad b_5 \approx -0.58\ldots \; , \qquad
b_6 \approx 2.3\ldots \times 10^{3} .
\end{equation}

\section{Discussion}
\setcounter{equation}{0}

\textbf{6.1} There are several other models of integrable QFT where the 
particle spectrum only consists of kink excitations with $S$-matrices 
related to the Boltzmann weights of the critical HHM. The most known among 
them is the RSOS(4) scattering theory arising from the perturbation of the 
(c=0.8) minimal CFT ${\cal M}_{5/6}$ by the operator $\phi_{1,3}$. The 
$S$-matrix of the general RSOS($p$) scattering theory, $p=3,4,\ldots$ , 
just coincide with the ``unitarised'' Boltzmann weights\footnote{The 
Boltzmann weights should also be supplied with simple gauge transformation 
factors which ensure the crossing-symmetry of the $S$-matrix.}
of the critical lattice RSOS model of Andrews, Baxter and Forrester (ABF)
\cite{abf} .
This is an interaction-round-a-face model on the square lattice where 
admissible neighbouring spin states belong to adjacent sites of the 
$A_p$ incidence diagram. The transfer matrix of this model commutes with 
the ``spin reversal'' operator generated by the reflection symmetry 
transformation of the incidence diagram and therefore it splits
into a direct sum of two transfer matrices acting diagonally in the even and
odd (with respect to the above symmetry) subspaces of the space of states. 
In the case $p=4$ each of these two transfer matrices coincide \cite{abf} 
with the transfer matrix of the hard hexagon model (modulo minor differences 
in boundary conditions which are irrelevant in the thermodynamic limit).
Therefore, for the purposes of the TBA calculations, the kink $S$-matrix of 
the RSOS($4$) model can be replaced by an appropriate ``hard hexagon'' type 
$S$-matrix which is given by the formulae (\ref{part1}),(\ref{eq:smat}) and 
(\ref{eq:normal}) with $\lambda$ and $F_{CDD}(\theta)$ replaced by
\begin{equation}
\mu = \frac{\pi}{5}, \qquad \lambda=-\frac{i}{5}, \qquad F_{CDD}(\theta)\equiv1,
\label{par1}
\end{equation}  
and all other quantities to remain unchanged. The calculation of Sect.~$2.4$
requires no modification for this case, one just needs to specialise the 
final results (\ref{eq:tbae}) and (\ref{eq:eofr}).

The ground state energy is 
\begin{equation}
E(R)={\rm (bulk\ term)}-\frac{m}{2 \pi} \int_{- \infty}^{\infty} \cosh 
\theta \, \log \big( 1+ e^{-\varepsilon_0(\theta)}\, \big)\,\, d \theta \ . 
\label{eq:eofr1}
\end{equation}
where the pseudo-energy $\varepsilon_{0} ( \theta )$ is determined by 
the following TBA integral equations 
\begin{equation}
\varepsilon_{j} ( \theta ) =\delta_{j,0}\, mR \, \cosh \theta  + 
\sum_{k=0}^{p-2} \int_{-\infty}^\infty \Phi_{j,k}(\theta-\theta') \log 
\big( 1+ e^{-\varepsilon_k( \theta')}\, \big)\,\, d \theta', \qquad 
j=0,1,\ldots,p-2.  \label{eq:tbae1}
\end{equation}
with $p=4$. The kernel $\Phi_{j,k}(\theta)$ is given by the same formula 
(\ref{eq:fnphi}) which, with account of (\ref{par1}), reads explicitly
\begin{equation}
\Phi_{j,k}(\theta) =-\,{1\over2\pi\cosh\theta}\,(\delta_{j,k+1}+\delta_{
j,k-1}) .
\end{equation}
Note that the sign factors $s_j$ which are present in
(\ref{eq:tbae}) do not appear in (\ref{eq:tbae1}) because the $\lambda$ in 
(\ref{par1}) has the opposite sign to the one in (\ref{eq:lambda}). 

The main technical point of our calculations was the diagonalization of
the transfer matrix (\ref{eq:trm}) by using the analytic Bethe-Ansatz solution 
\cite{bazresh} for the HHM. In fact, the aforementioned paper contains 
the results of this diagonalisation procedure for the complete hierarchy of 
ABF RSOS lattice models \cite{abf} (and their higher fused generalisations 
\cite{jimbo}). The calculations of the present paper are also easily 
extended to the case when the kink $S$-matrices are expressed through the 
Boltzmann weights of these more general models \cite{Hollowood}. 
In particular, for the general 
RSOS($p$) scattering theories one gets in this way 
the TBA equations (\ref{eq:tbae1}), which are identical to those 
conjectured by Zamolodchikov \cite{zam2}.
Note also, that essentially the same set of TBA equations (or more precisely 
their lattice counterparts, which differ from (\ref{eq:tbae1}) only in the 
form of the energy term $mR\cosh\theta$) arose previously
\cite{bazresh} in the analysis
of finite size corrections to the critical ABF RSOS models.

\textbf{6.2} Another integrable QFT where the scattering theory is related 
to the HHM is obtained from the $c=-3/5$ minimal CFT ${\cal M}_{3/5}$ 
perturbed by the operator $\phi_{2,1}$ with dimension $\Delta=3/4$. This 
QFT is also covered by Smirnov's conjecture \cite{smirnov}. It was 
specialised to this case in \cite{mus2}, resulting in a kink $S$-matrix 
that can be written exactly as in (2.10) and (2.12) with
\begin{equation}
\mu={3\pi\over5}, \qquad \lambda=-{3i\over5},\qquad
F_{CDD}(\theta)={\sinh\theta+i\sin(\pi/3)\over \sinh\theta-i\sin(\pi/3)} . 
\label{m35}
\end{equation}
Evaluating the integral in (\ref{part}) for $\mu=3\pi/5$  
and substituting the result in the normalisation factor 
(\ref{eq:normal}) one gets explicitly
\begin{equation}
R(\theta) = \frac{\sin \left( \frac{2 \pi}{5} - \frac{3 i \theta}{5} \right)
\sin \left( \frac{\pi}{5} - \frac{3 i \theta}{5} \right)}{\sin \left( 
\frac{2 \pi}{5} + \frac{3 i \theta}{5} \right) \sin \left( \frac{\pi}{5} +
\frac{3 i \theta}{5} \right)} .
\end{equation}

The calculations of Sect.~3 and Sect.~4 can be repeated in this case with
certain modifications. The main difference is related to the fact 
that for $\mu=3\pi/5$ the string structure of the solutions of the BAE 
equations (\ref{eq:nbae}) changes. Numerical calculations suggest that in 
this case only 
1-strings in (3.2) and the shifted 1-strings (3.3) are thermodynamically 
significant. We believe this statement is correct 
 and claim it as a conjecture. 
The rest of the calculations closely parallel those given in the case 
$\mu=\pi/5$ and requires no further assumptions. As a result one obtains that 
the ground state energy $E(R)$ is still given by the expression 
(\ref{eq:eofr1}) while the pseudo-energy $\varepsilon_0(\theta)$ is 
determined by the following TBA equations
\begin{eqnarray}
\varepsilon_0(\theta) & = & mR\cosh\theta+\int_{-\infty}^\infty
\Phi(\theta-\theta')\left(\log\big( 1+ e^{-\varepsilon_0( \theta')}\, \big)- 
\log\big( 1+ e^{-\varepsilon_1( \theta')}\, \big)\right)\,\, d \theta',  
\nonumber \\
\varepsilon_1(\theta) & = & \phantom{mR\cosh\theta} + \int_{-\infty}^\infty
\Phi(\theta-\theta') \left(\log\big( 1+ e^{-\varepsilon_1( \theta')}\, \big)-
\log\big( 1+ e^{-\varepsilon_0( \theta')}\, \big)\right)\,\, d \theta',  
\label{eq:m35tbae}
\end{eqnarray}
where the kernel is given by
\begin{equation}
\Phi(\theta)={\sqrt{3}\over \pi}{\sinh(2\theta)\over\sinh(3\theta)}
\end{equation}
Note also that 
\begin{equation}
\Phi(\theta)={i\over2\pi}\partial_\theta \log F_{CDD}(\theta)
\end{equation}
where $F_{CDD}$ is the CDD factor defined in (\ref{m35}).

Apparently the same TBA equations for this QFT were
conjectured in  \cite{rav1}\footnote{In \cite{rav1} the TBA equations for 
the $\mathcal{M}_{3/5}$ minimal CFT perturbed by the operator $\phi_{2,1}$ 
were obtained by ``orbifolding'' of the TBA equations associated with the 
product of $A_2\otimes A_2$ Dynkin diagrams. The latter were previously 
conjectured in \cite{rav2,rav3} (in a rather general form) as field theory 
counterparts for the TBA equations of lattice RSOS models connected with the 
simply laced algebras obtained in ref. \cite{bazresh2}. 
It should be noted that the relevant equations in refs. \cite{rav1,rav2,rav3}
contain some consistent misprints.  However, tracing their
derivations back to the original reference \cite{bazresh2} one can recover
the TBA equations (\ref{eq:m35tbae}) exactly as given here. We thank 
F. Ravanini and R. Tateo for explaining this issue.} 
from different arguments. 
 
\textbf{6.3} Yet another related QFT is obtained as the perturbation  
of the minimal CFT $\mathcal{M}_{5/6}$ by the operator $\phi_{1,2}$ of
dimension $\Delta = \frac{1}{8}$. Again the $S$-matrix of the fundamental kink
states (as given by Smirnov's conjecture \cite{smirnov}) is expressed through
the Boltzmann weights of the HHM ($2.3$-$2.7$) with $\mu=\pi/5$. However, 
unlike all of the above examples, the bound state state structure of this 
QFT appears to be extremely complicated. Apart from the fundamental kinks
it contains higher kinks and breathers as bound states of the fundamental 
kinks. These higher kinks and breathers produce even more new kink and 
breather states and so on. A few of the lower kink and breather states were 
found in \cite{koubek}. It would be interesting to see whether this bootstrap 
procedure closes at all. Particularly, some preliminary estimates 
show that the spectrum of this theory contains at least a dozen particles.
In any case all of the kink-kink $S$-matrices are expressed in terms of the
HHM Boltzmann weights ($2.3$-$2.7$) and therefore the TBA calculation of the 
present paper could be generalised to this case provided the  bootstrap 
program is completed.

\vspace{1cm}
The above considerations were restricted only to the calculation 
of the ground state energy in the finite-size IQFT's by the TBA method.
Recently,  a few  new approaches were introduced \cite{blz1,blz2,blz3,f1,d}
which allow one 
to calculate also excited state energies. In ref. \cite{blz3} it was 
shown that the functional equations for the IQFT ``commuting
transfer-matrices'' introduced in \cite{blz1} can be transformed \cite{kp} to 
the integral equations which generalise the equations of TBA to the
excited states. The results of \cite{blz3} apply to massive IQFT's obtained by 
$\phi_{1,3}$ perturbations of the minimal CFT's which are related to the 
$U_q(\widehat{sl}(2))$ quantum algebra.   
A suitable generalisation of this approach to treat the $\phi_{1,2}$ 
or $\phi_{2,1}$ perturbations of the minimal CFT's (in particular, the PTIM
considered above) should be associated with the $q$-deformed twisted Kac-Moody
algebra $A^{(2)}_2$. Some results in this direction were obtained in \cite{f2}.
We hope to address this problem in future publications.


\section*{Acknowledgements}

The authors are grateful to G. Mussardo for very helpful
discussions, especially regarding the use of the TCSA program STRIP 
\cite{strip}. V. B. thanks A. B. Zamolodchikov and S. O. Warnaar 
for interesting discussions.


\end{document}